\documentclass[12pt,a4paper]{ws-ijmpa}
\usepackage[super,compress]{cite}
\usepackage{graphicx,amsmath}
\usepackage{multicol}
\usepackage{subcaption}

\newcommand{\bea}{\begin{eqnarray}}
\newcommand{\eea}{\end{eqnarray}}
\newcommand{\nn}{\nonumber}

\begin{document}	

\title{Exploring new physics via effective interactions}

\author{Ekata Singh}

\address{Dr. RML Govt. Degree College, Aonla (MJP Rohilkhand University), Bareilly, Uttar Pradesh, India}

\author{Hrishabh Bharadwaj}

\address{Rajkiya Mahila Mahavidyalaya, Budaun - 243601 (MJP Rohilkhand University, Bareilly), Uttar Pradesh, India\\
hrishabhphysics@gmail.com}

\author{Devisharan}
\address{Dr. RML Govt. Degree College, Aonla (MJP Rohilkhand University), Bareilly, Uttar Pradesh, India}

\maketitle

\begin{history}
\received{Day Month Year}
\revised{Day Month Year}
\end{history}

\begin{abstract}
We investigate self-conjugate Dark Matter (DM) particles that primarily interact with standard model electroweak gauge bosons within an effective field theoretical framework. Our analysis focuses on effective contact interactions, invariant under the standard model gauge group, between Majorana fermions, and real scalar DM, with SM neutral electroweak gauge bosons. We calculate the Wilson coefficients for interaction terms up to dimension-8 and establish constraints on the theory's parameters. These constraints are derived from the observed relic density, and indirect detection observations. We discuss the potential for dark matter-nucleon scattering in direct detection experiments. Additionally, we utilize low-energy LEP data to assess sensitivity to the pair production of low-mass ($\leq$ 80 GeV) DM particles. Furthermore, we highlight the potential of the proposed International Linear Collider (ILC) in probing effective operators through the pair production of DM particles with masses $\geq$ 50 GeV in association with mono-photons.

\keywords{Indirect detection; nuclear recoil; dark matter.}
\end{abstract}

\ccode{PACS numbers:}


\section{Introduction}

Dark matter, constituting approximately 23\% of the Universe's energy density and about 75\% of its total matter, has been inferred from various cosmological and astrophysical observations. The Planck Collaboration has provided precise measurements of DM density, with a relic density value of $\Omega_{DM} h^2 = 0.1198 \pm 0.0012$ \cite{Planck:2018vyg}. Despite these advancements, the nature of DM remains elusive.

Direct detection experiments like CRESST \cite{CRESST:2016qpj}, DAMA/LIBRA \cite{Bernabei:2013xsa, Bernabei:2018yyw}, XENON100 \cite{XENON100:2016sjq,XENON:2017lvq}, CoGeNT \cite{CoGeNT:2012sne}, LUX \cite{LUX:2016ggv}, CDMS \cite{CDMS:2013juh}, and PandaX-II \cite{PandaX-II:2017hlx} aim to measure the recoil momentum of scattered atoms or nucleons by DM in detectors. Collider experiments, both present and proposed, focus on identifying DM particle production through mono or di-jet events with missing energy. Indirect experiments, such as FermiLAT \cite{Fermi-LAT:2015att}, HESS \cite{HESS:2013rld}, and AMS-02 \cite{AMS:2014xys,AMS:2016oqu } seek evidence of excess cosmic rays resulting from DM annihilation to Standard Model particles.

Effective Field Theory (EFT) provides a model-independent approach, treating DM-SM interactions as contact interactions described by non-renormalizable operators \cite{Chen:2013gya,Chae:2012bq}. This framework allows the study of various DM phenomenological aspects, exploring interactions through different processes like annihilation and scattering. Sensitivity analyses for DM-quark and DM-gauge Boson effective interactions have been conducted at the LHC in a model-independent manner, aiming to constrain the parameter space of simplified models and popular scenarios \cite{Cotta:2012nj,Crivellin:2015wva,Chen:2015tia,Bell:2012rg}.

In the realm of deep inelastic lepton-hadron scattering, analyses have been carried out on twist-2 operators and their renormalization-group equations, providing insights into DM-nucleon scattering induced by quark operators. Additionally, studies on one-loop effects in DM-nucleon scattering induced by twist-2 quark and gluonic operators have been conducted \cite{Drees:1993bu,Hisano:2010ct,Hisano:2010yh,Hisano:2017jmz}. Overall, research efforts continue to unravel the elusive nature of dark matter through a combination of direct and indirect detection experiments, collider searches, via effective interactions.

In this study, we examine dark matter (DM) currents primarily interacting with the Standard Model (SM) electroweak neutral gauge bosons through gauge-invariant effective operators within the $SU(2)_L \times U(1)_Y$ framework. To uphold the gauge symmetry of the SM at all energy scales, we confine our dark matter candidates to self-conjugate entities: a Majorana fermion and a real scalar  all of which are SM gauge singlets. We concentrate on operators involved in the non-relativistic, spin-independent scattering interaction between dark matter and nucleons. We exclude operators that are suppressed by the velocity of either dark matter or nucleons. Section 2 details the formulation of the effective interaction Lagrangian for fermionic, scalar, and vector DM with SM electroweak neutral gauge bosons using higher dimensional operators.

In Section 3, we constrain the coefficients of the effective Lagrangian based on the observed relic density and conduct a consistency check through indirect and direct experiments. The constraints from the Large Electron-Positron Collider (LEP) and the sensitivity analysis of the coefficients for the effective operators at the proposed International Linear Collider (ILC) are discussed in Section 4. Our findings are summarized in Section 5.

\section{Model}
\label{model}
\par The Standard Model (SM) is augmented by a set of SM gauge-invariant operators with dimensions up to 8. The effective Lagrangian governing the interaction between a real scalar ($\phi$) and Majorana fermion ($\psi$) dark matter with the SM neutral electroweak gauge bosons is provided as follows: 

\bea
{\mathcal L}_\psi &=& \frac{{\alpha^{\psi}_{S}}}{\Lambda^4}\ {O}_{S}^{\psi}+\frac{{\alpha^{\psi}_{P}}}{\Lambda^4}\ {O}_{P}^{\psi}+\frac{{\alpha^{\psi}_{T}}}{\Lambda^4}\ {O}_{T}^{\psi},\label{LFDM}\\
{\cal L}_\phi &=&\frac{{\alpha^{\phi}_{S}}}{\Lambda^4}\ {O}_{S}^{\phi}+ \frac{{\alpha^{\phi}_{T}}}{\Lambda^4}\ {O}_{T}^{\phi}, \label{LSDM}
\eea

where $\Lambda$ denotes the effective theory's cut-off scale and $\alpha_i$ is the respective strength of the effective interaction. The effective operators for fermionic and scalar DM are given as:
\bea
 {O}^{\psi}_S&=&m_{\psi}\ \overline{\psi}\ \psi\ B_{\mu\nu}\ B^{\mu\nu},\nn\\
 {O}^{\psi}_P&=&m_{\psi}\ \overline{\psi}\ i \gamma^5\ \psi\ B_{\mu\nu}\ \widetilde{B}^{\mu\nu}, \nn\\ 
 { O}^{\psi}_{T}&=& 
\overline{\psi}\ i\ {\partial}^\mu\ {\gamma}^\nu\ \psi \ O_{\mu\nu} \ +\ \text{h.c.}, \nn\\
 {O}^{\phi}_S&=&  m_{\phi}^2\ \phi^2\ B_{\mu\nu}\ B^{\mu\nu}, \nn\\
{O}^{\phi}_{T}&=& \phi\ i\ \partial^\mu\ i\ \partial^\nu\ \phi\ O_{\mu\nu}\ +\ \text{h.c.},
\label{Operators}
\end{eqnarray}

with $O_{\mu\nu}\ =\ B^\alpha_\mu\ B_{\alpha \nu}\ -\ \frac{1}{4}\ g_{\mu \nu}\ B^{\alpha \beta}\ B_{\alpha \beta}$, where $B_{\mu\nu}$ represents the field strength tensor associated with the SM $U(1)_Y$ gauge. For fermionic dark matter, such operators may emerge from interactions of the form $\overline{\psi}\ \gamma^\mu \left(a + b\ \gamma_5 \right) \chi\ V_\mu$, where $\psi$ represents the dark matter candidate, $V_\mu$ denotes the Standard Model electroweak gauge boson, and $\chi$ is the beyond SM fermion originating from theories like supersymmetry \cite{Nilles:1983ge} or extra dimensions \cite{Appelquist:2000nn}.

\section{DM phenomenology}

\subsection{Relic density}
\par In the early Universe, dark matter (DM) particles interacted with the plasma and maintained thermal equilibrium through the annihilation and creation of DM particles.  However, as the Universe expanded and they became non-relativistic, they departed from thermal equilibrium. Eventually, they got `frozen-out' and became a cold relic when their annihilation rate fell below the Hubble expansion rate. The evolution process is governed by the Boltzmann equation, and the final expression for DM relic density can be approximated to \cite{Bauer:2017qwy,Dodelson:2003ft}:

\bea
\Omega_{\psi/\phi}\ h^2 &\approx& 0.12\ \sqrt{\frac{g_\star(x_f)}{100}}\ \left(\frac{x_f}{28}\right)\ \frac{2\times 10^{-26}\ cm^3/s}{\langle\sigma\	v\rangle},
\eea
where $x_f,\ g_\star(x_f),\ \langle \sigma\ v\rangle$ are the ratio of DM mass and temperature at freeze-out, effective degrees of freedom at freeze-out and thermally averaged DM annihilation cross-section respectively. The thermally averaged cross-section can be written as $\langle \sigma\ v \rangle\ =\ a\ +\ b\ \langle v^2 \rangle\ +\ c\ \langle v^4 \rangle +\mathcal{O}(v^6)$, where $\langle v^2 \rangle\ =\ \frac{6}{x_f}\ \&\ \langle v^4 \rangle\ =\ \frac{60}{x_f^2}$. The thermally averaged cross-sections for the processes contributing to DM relic density are given in \ref{App}.

\begin{figure}[ht]
\centering

\begin{subfigure}{0.45\textwidth}
  \centering
  \includegraphics[width=\linewidth]{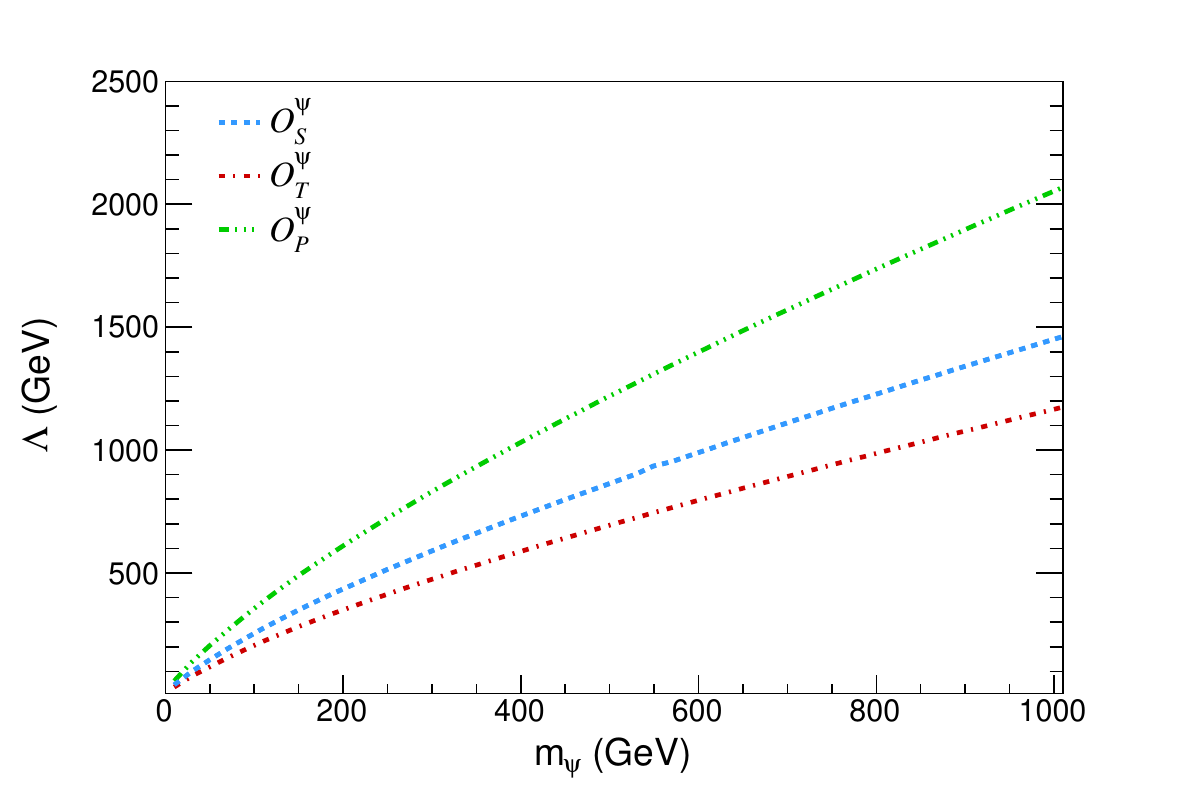}
  \caption{}
  \label{fig1}
\end{subfigure}
\hfill
\begin{subfigure}{0.45\textwidth}
  \centering
  \includegraphics[width=\linewidth]{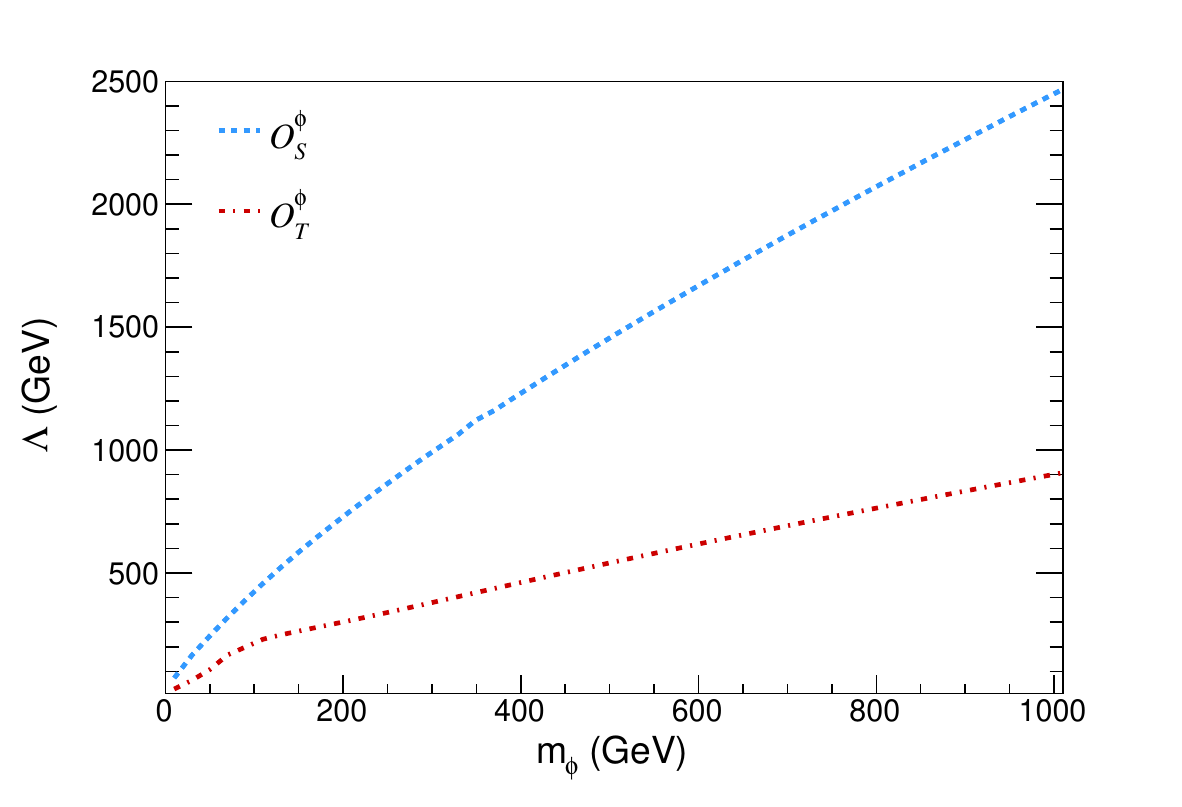}
  \caption{}
  \label{fig2}
\end{subfigure}

\caption{The contour lines corresponding to DM relic density $\Omega\ h^2 = 0.1198$ for respective (a) fermionic, and (b) scalar DM operators. The parameter space below the contour line is allowed from the relic density observations.}
\label{fig:relic}
\end{figure}

\par We have used FeynRules \cite{Alloul:2013bka} to generate the interaction vertices and other necessary model files. These model files were subsequently employed in the MadDM \cite{MadDM} package for calculating the relic density. Contour plots in the $\Lambda- m_{\psi/\phi}$ plane, satisfying the current DM relic density of $\Omega\ h^2\ =\ 0.1198$ \cite{Planck:2018vyg} for DM mass ranging from 10 GeV to 1 TeV, are depicted in Figures \ref{fig1} and \ref{fig2}. The regions below the contour lines are allowed from relic density constraints. Throughout the relic density computation, the respective coupling strengths $\alpha_i$ are taken to be 1.

\subsection{Indirect detection}
The annihilation of dark matter in the densely populated areas of the Universe would result in a substantial influx of high-energy SM particles. The High Energy Stereoscopic System (H.E.S.S.) \cite{HESS:2018zix} utilizes indirect detection techniques to investigate the presence of dark matter. By observing high-energy gamma rays emitted from specific regions of interest, such as the galactic center or dwarf galaxies, H.E.S.S. endeavors to discern potential signals arising from dark matter annihilation or decay. These processes may give rise to gamma rays as secondary particles when dark matter particles interact and transform into standard model particles. Through precise measurements of gamma-ray fluxes and energy spectra, H.E.S.S. furnishes valuable insights into the characteristics and distribution of dark matter in the universe, thereby placing constraints on its properties. 
  \par The expressions for the cross-section of dark matter annihilation into a photon pair are provided as follows:
\begin{eqnarray}
\sigma^{\psi}_{S}\ v\ \left(\psi\ \overline{\psi}\to \gamma \gamma\right)&\simeq&
 \frac{2}{\pi}\ 
\frac{{\alpha^{\psi}_S}^2}{\Lambda^8}\ \cos^4\theta_w\ m_{\psi}^6\ v^2\\
\label{indFS}
\sigma^{\psi}_{P}\  v\ \left(\psi\ \overline{\psi}\to \gamma \gamma\right)&\simeq&
 \frac{2}{\pi}\ 
\frac{{\alpha^{\psi}_P}^2}{\Lambda^8}\ \cos^4\theta_w\ m_{\psi}^6\  \left(1- \frac{ v^2}{4}\right)\\
\label{indFP}
 \sigma^{\psi}_{T}\  v\ \left(\psi\ \overline{\psi}\to \gamma \gamma\right)&\simeq&
 \frac{2}{3 \pi}\ 
\frac{{\alpha^{\psi}_T}^2}{\Lambda^8}\ \cos^4\theta_w\ m_{\psi}^6\ v^2\\
\label{indFT}
 \sigma^{\phi}_{S}\ v\ \left(\phi\ {\phi}\to \gamma \gamma\right)&\simeq&
 \frac{16}{\pi}\ 
\frac{{\alpha^{\phi}_S}^2}{\Lambda^8}\ \cos^4\theta_w\ m_{\phi}^6\  \left(1- \frac{ v^4}{4}\right)\\
\label{indSS}
 \sigma^{\phi}_{T}\  v\ \left(\phi\ {\phi}\to \gamma \gamma\right)&\simeq&
 \frac{4}{15 \pi}\ 
\frac{{\alpha^{\phi}_T}^2}{\Lambda^8}\ \cos^4\theta_w\ m_{\phi}^6\ v^4
\label{indST}
\end{eqnarray}
\begin{figure}[ht]
\centering

\begin{subfigure}{0.45\textwidth}
  \centering
  \includegraphics[width=\linewidth]{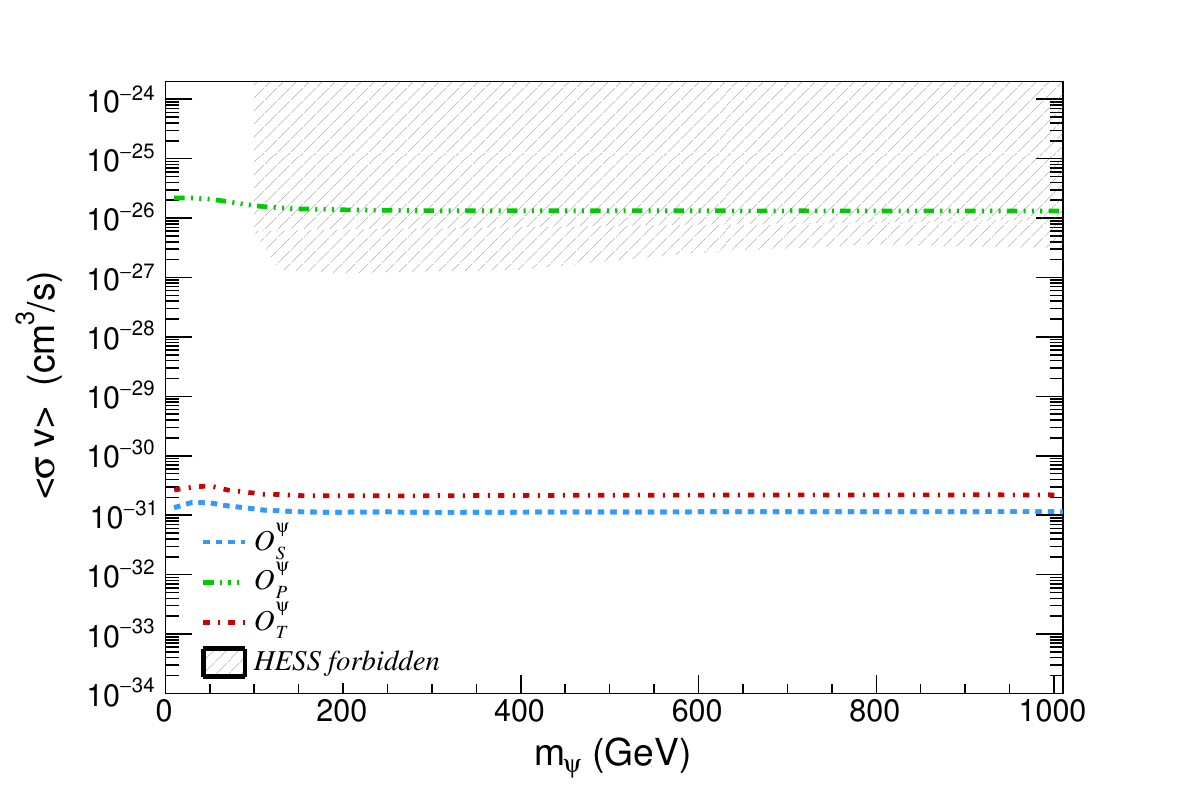}
  \caption{}
  \label{fig3}
\end{subfigure}
\hfill
\begin{subfigure}{0.45\textwidth}
  \centering
  \includegraphics[width=\linewidth]{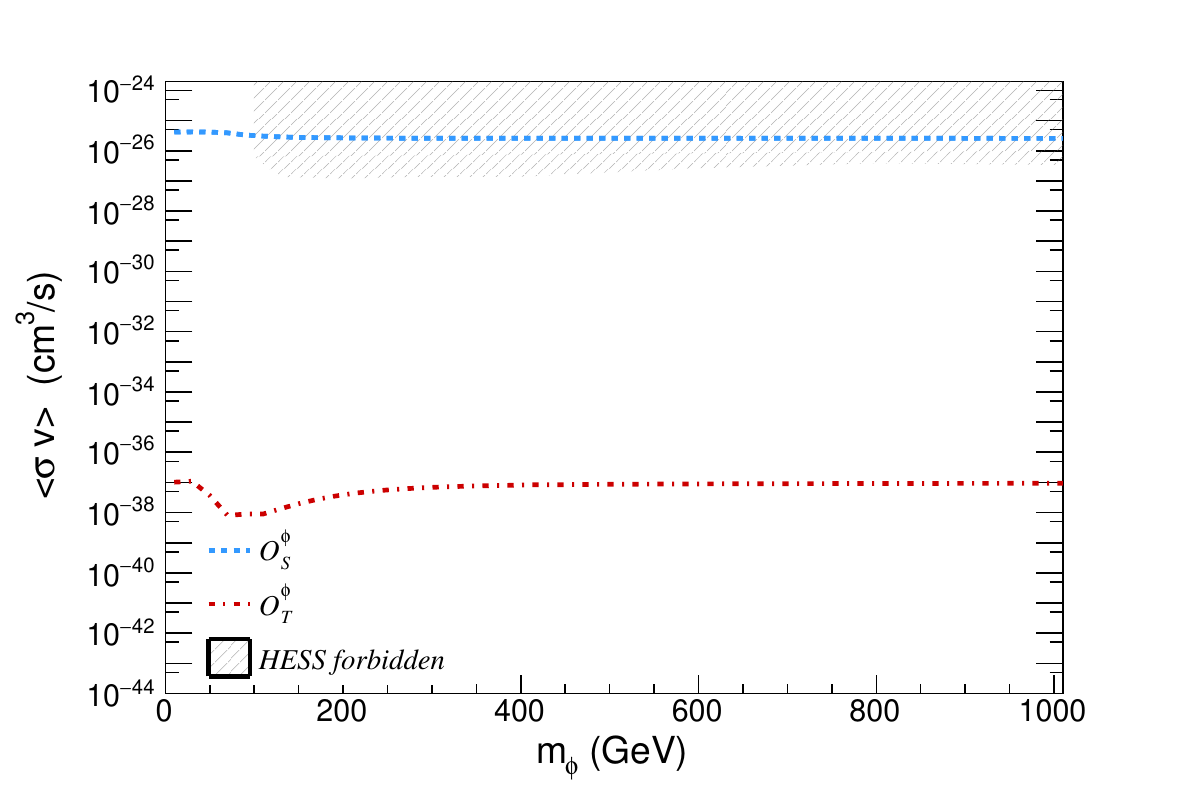}
  \caption{}
  \label{fig4}
\end{subfigure}

\caption{Variation of the (a) fermionic, and (b) scalar DM annihilation cross-section into $\gamma\ \gamma$ for each operator, depicted by colored lines, as a function of the DM mass, while keeping all other parameters consistent with the observed relic density. The shaded region is excluded based on data from H.E.S.S.}
\label{fig:ind}
\end{figure}
We explore the mass range of DM from 10 GeV to 1 TeV. In the thermal averaged cross-sections expressed in equations \eqref{indFS}-\eqref{indST}, the terms that do not depend on the DM relative velocity $v$, are solely produced by s-wave annihilation. Whereas, terms that are proportional to $v^2$ are associated with p-wave annihilation cross-sections. we have taken DM relative velocity to be $\sim 10^{-3}\ c.$ The equations given in equations \eqref{indFS}-\eqref{indST} reveal that $\mathcal{O}^\psi_P$ is $s-$wave and $\mathcal{O}^\psi_S\ \&\ \mathcal{O}^\psi_T$ are $p-$wave suppressed for fermionic DM. Where as $\mathcal{O}^\phi_S$ is $s-$wave and $\mathcal{O}^\phi_T$ is $d-$wave suppressed. In computing DM annihilation cross-section to photon pair, we have used parameters allowed from relic density constraints for a given DM mass $m_{\psi/\phi}$.  For a given $m_{\psi/\phi}$, the respective annihilation cross-section $\langle \sigma\ v \rangle$ is shown in figures \ref{fig3} and \ref{fig4}. The region above the respective line, is allowed, the shaded region corresponds to the H.E.S.S. indirect detection forbidden region.

\subsection{Nuclear Recoil}

\begin{figure}[h]
\centerline{\includegraphics[scale=0.9]{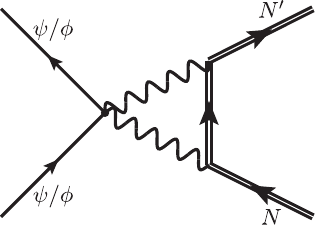}}
\caption{Feynman diagram representing the scattering of DM  particles ($\psi/ \phi$) with the nucleons.}
\label{DDloop}
\end{figure}

Experiments focusing on direct detection \cite{XENON:2017lvq,Bernabei:2013xsa,CoGeNT:2012sne,CRESST:2016qpj,XENON100:2016sjq,LUX:2016ggv,Bernabei:2018yyw,PandaX-II:2017hlx,CDMS:2013juh} aim to observe the scattering of DM particles with nucleons or atoms. These experiments are specifically designed to measure the recoil momentum of the nucleons or atoms within the detector material. Such scattering events are broadly categorized into (i) DM-nucleon, (ii) DM-atom, and (iii) DM-electron interactions. In our model, where DM does not exhibit a direct interaction with leptons, quarks or gluons at the tree level, these interactions can only occur at loop levels, as illustrated for DM nucleon scattering in figure \ref{DDloop}, which constitutes the next focus of the ongoing research work.

\section{Collider searches}
\subsection{Constraints from LEP}
Utilizing existing findings and observations from LEP data, constraints on effective operators can be established. The cross-section for the process $e^+e^-\to \gamma^\star + \ \psi\ \overline{\psi}\ (\phi\ \phi)$ is compared with the joint investigation performed by DELPHI and L3 collaborations for $e^+e^-\ \to\ \gamma^{\star} + Z \to q\ \bar q + \nu_{l}\bar\nu_{l}$ at a \emph{c.m.} energy $\sqrt{s}$ of $196.9$ GeV and an integrated luminosity of 0.679 fb$^{-1}$. Here, $q$ denotes the light quarks, namely $u,\ d\ \&\ s$, and $\nu_{l}$ represents Standard Model neutrinos, $\nu_e,\,\nu_{\mu},\nu_\tau$.

\begin{figure}[h]
\centerline{\includegraphics[scale=0.9]{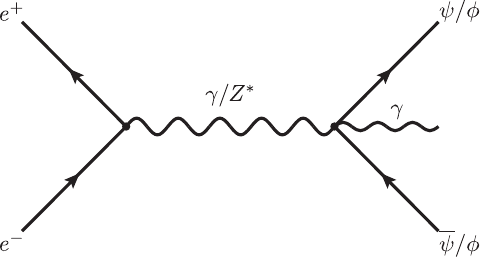}}
\caption{Feynman diagram for the production of a dark matter pair with a photon in an $e^+ e^-$ collider.}
\label{FeynD}
\end{figure}

\begin{figure}[ht]
\centering

\begin{subfigure}{0.45\textwidth}
  \centering
  \includegraphics[width=\linewidth]{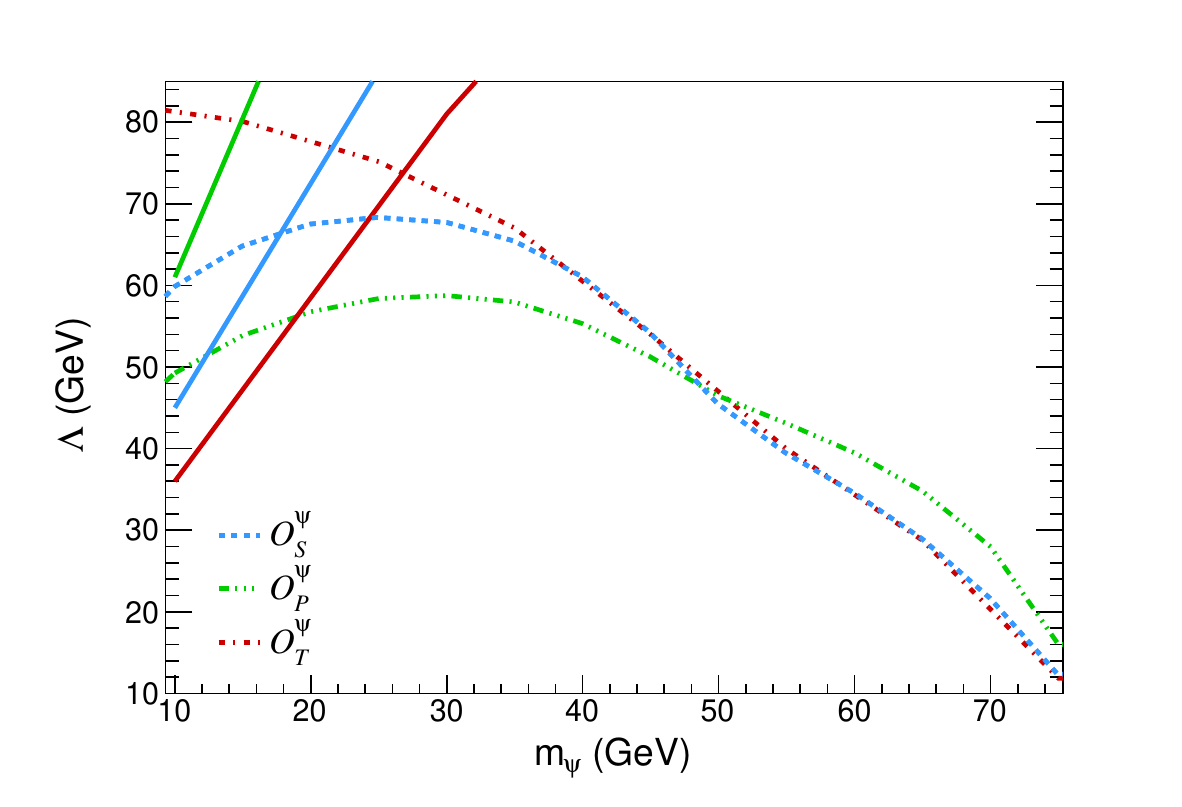}
  \caption{}
  \label{LEP1}
\end{subfigure}
\hfill
\begin{subfigure}{0.45\textwidth}
  \centering
  \includegraphics[width=\linewidth]{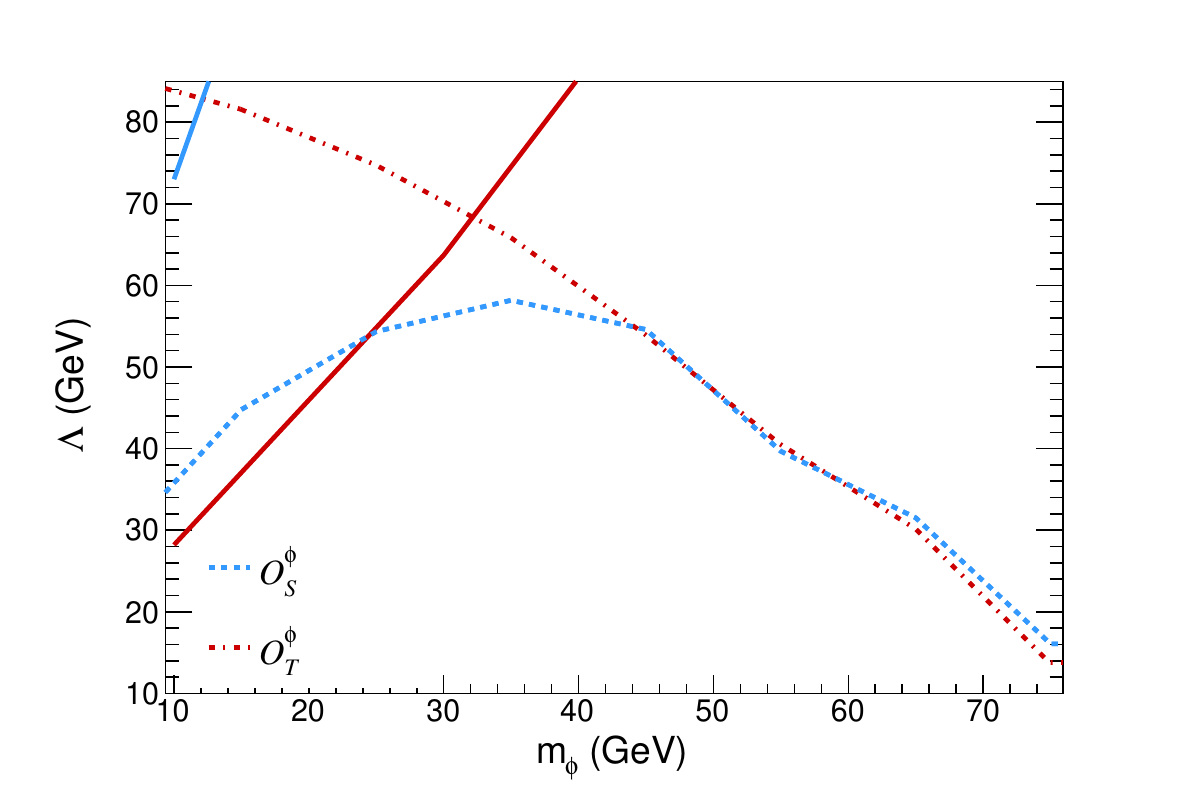}
  \caption{}
  \label{LEP2}
\end{subfigure}

\caption{Dashed lines illustrate the contours within the parameter space defined by the fermionic DM mass and the kinematic reach for (a) $e^+e^- \to  \psi\ \bar{\psi} + \gamma^\star \to \,\,\not\!\!\!E_T + q_{i}\bar q_{i}$, and (b) $e^+e^- \to \phi\ \phi + \gamma^\star \to \,\,\not\!\!\!E_T + q_{i}\bar q_{i}$ at a center-of-mass energy of $\sqrt{s}$ = 196.9 GeV, considering an integrated luminosity of 679.4 pb$^{-1}$. These contours adhere to the constraint $\delta\sigma_{\rm tot}$ = 0.032 pb, derived from the combined analysis of DELPHI and L3 \cite{ALEPH:2013dgf}. The region below the dashed lines is excluded based on LEP observations. Additionally, the regions below the solid lines corresponding to specific operators satisfy the relic density constraint $\Omega_{\rm DM}h^2 \leq 0.1198$.}
\label{fig:LEP}
\end{figure}

The Feynman diagrams that contribute to the producion of a photon $\gamma$ with missing energy, resulting from effective operators as outlined in equation \eqref{Operators}, at the $e^-,e^+$ lepton collider are illustrated in figure \ref{FeynD}. The measured cross-section for this process is determined to be $0.055$ pb, accompanied by the statistical error $\Delta\sigma_{\rm stat.}$, systematic error $\Delta\sigma_{\rm sys.}$, and total error $\Delta\sigma_{\rm total}$ of $0.031$ pb, $0.008$ pb, and $0.032$ pb, respectively, as reported in \cite{ALEPH:2013dgf}. Consequently, the contribution arising from an additional channel involving final states with dark matter pairs, leading to missing energy plus two quark jets, can be constrained based on the observed $\Delta\sigma_{\rm total}$. In figures \ref{LEP1} and \ref{LEP2}, we have depicted 95\% confidence level dashed line contours that satisfy $\Delta\sigma_{\rm total} \simeq 0.032$ pb, corresponding to the operators in the dark matter mass-$\Lambda$ plane. The region below the dashed lines, as indicated, is not allowed by the combined analysis of LEP. Whereas the solid lines correspond to the relic density contours.

\subsection{Constraints from ILC}

We now investigate the processes involving DM pair production along with an on-shell photon at the proposed International Linear Collider (ILC) for the DM mass spectrum ranging from approximately 10 to 250 GeV. The specific processes under consideration are: $e^+\,e^-\,\rightarrow\ \psi\ \overline{\psi}\ \gamma $, and $e^+\,e^-\,\rightarrow\ \phi\ \phi\ \gamma $, depicted in figure \ref{FeynD}. The primary SM background for $e^+e^-\to\ \not \!\! E_T\ \gamma$ process arises from the $e^+\,e^-\ \rightarrow\ Z\ \gamma \to  \nu\,\overline{\nu}\ \gamma$ process.

\begin{table}[h]
\centering
\begin{tabular}{ccc} \toprule
 & $\textit{ILC-250}$ & $\textit{ILC-500}$ \\ \colrule
$\sqrt{s}\ \left( \textit{GeV}\right )$ & 250 & 500 \\
$L_{int}\ \left(\textit{$fb^{-1}$}\right)$ & 250 & 00 \\
$\sigma_{BG}\ \left(\textit{pb}\right)$ & 1.07 & 1.48\\  \botrule
\end{tabular}
\caption{Accelerator parameters}
\label{table:accelparam}
\end{table}

The examination of background and signal processes with the accelerator parameters outlined in the technical design report for ILC \cite{Behnke:2013lya, Behnke:2013xla}, as presented in Table \ref{table:accelparam}, involves simulation of SM background and DM signal. Model files are created using FeynRules \cite{Alloul:2013bka}, and events are generated with Madgraph \cite{Alwall:2014hca}. For event selection and analysis, MadAnalysis 5 \cite{Conte:2012fm} has been utilized.

In addition to the basic selection criteria, which include cuts on the transverse momentum of the photon ($p_{T_{\gamma}} \geq$ 10 GeV) and the pseudo-rapidity of the photon ($\left\vert\eta_\gamma\right\vert\leq$ 2.5), a selection requirement related to the photon energy in the context of on-shell production is applied. Specifically, events are excluded if the following condition is met:

\bea
\frac{2\,E_\gamma}{\sqrt{s}} = 1\ -\ \frac{(m_Z^2\ \pm\ 10\ m_Z \Gamma_Z)}{s}
\eea

This translates to the rejection of events where $ 2\,E_\gamma/\sqrt{s} \in \left[0.8,0.9\right]$ and $\left[0.95,0.98\right]$ for center-of-mass energies ($\sqrt{s}$) of 250 GeV and 500 GeV, respectively.

\begin{figure}[ht]
\centering

\begin{subfigure}{0.32\textwidth}
  \centering
  \includegraphics[width=\linewidth]{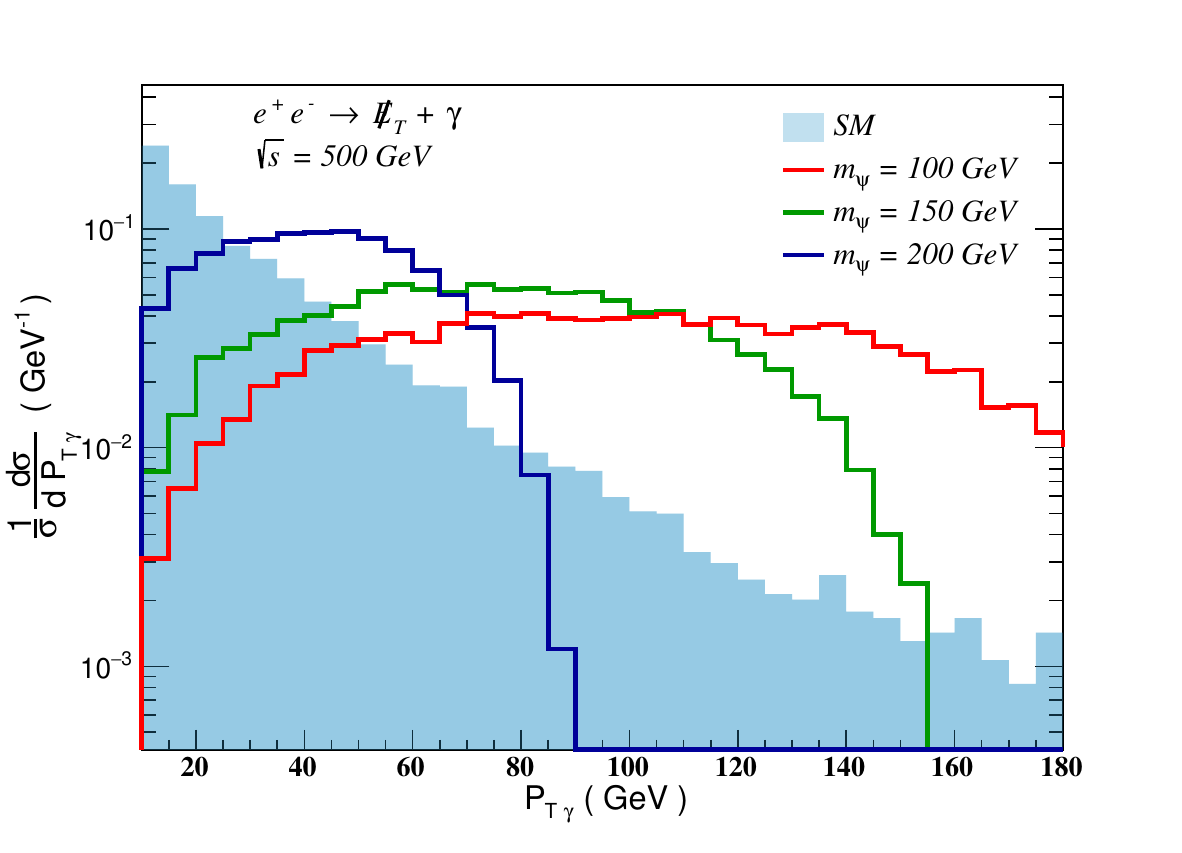}
  \caption{}
  \label{DBfig1}
\end{subfigure}
\hfill
\begin{subfigure}{0.32\textwidth}
  \centering
  \includegraphics[width=\linewidth]{FS_pt.pdf}
  \caption{}
  \label{DBfig2}
\end{subfigure}
\begin{subfigure}{0.32\textwidth}
  \centering
  \includegraphics[width=\linewidth]{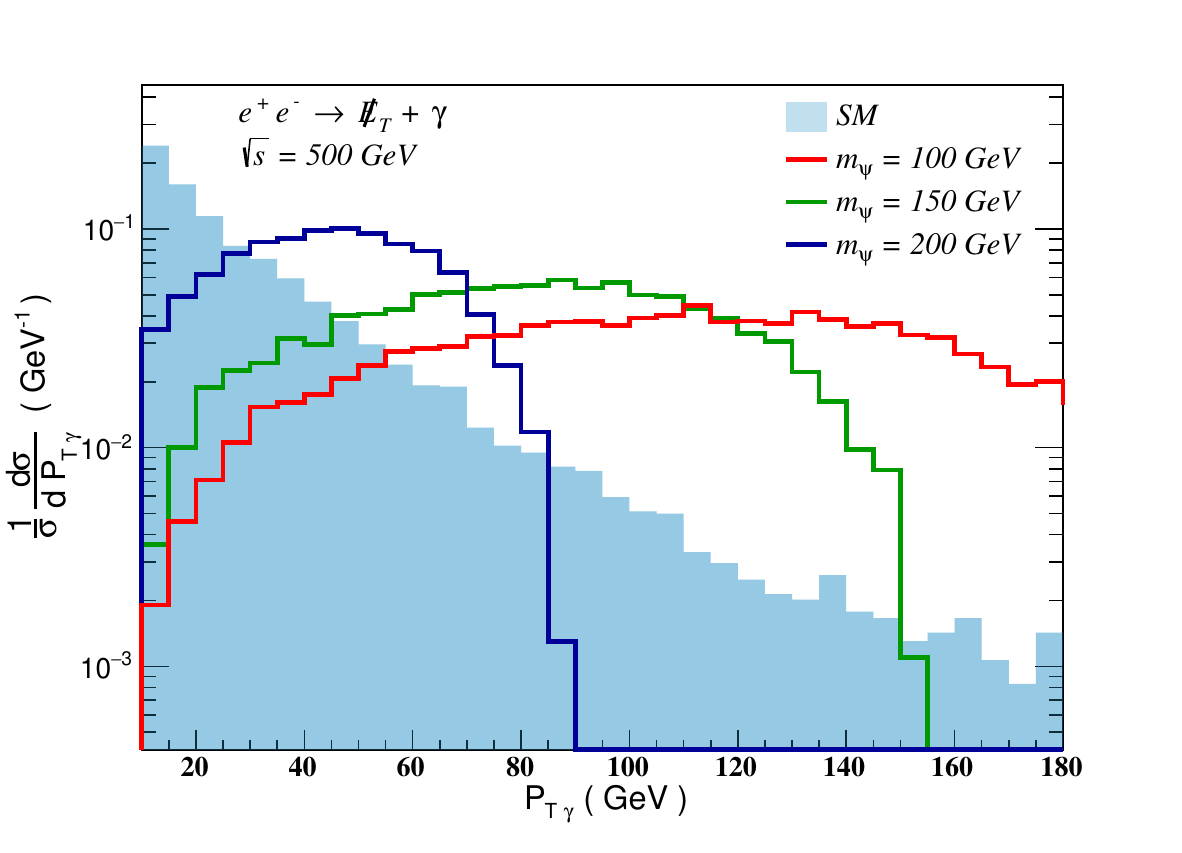}
  \caption{}
  \label{DBfig3}
\end{subfigure}
\begin{subfigure}{0.32\textwidth}
  \centering
  \includegraphics[width=\linewidth]{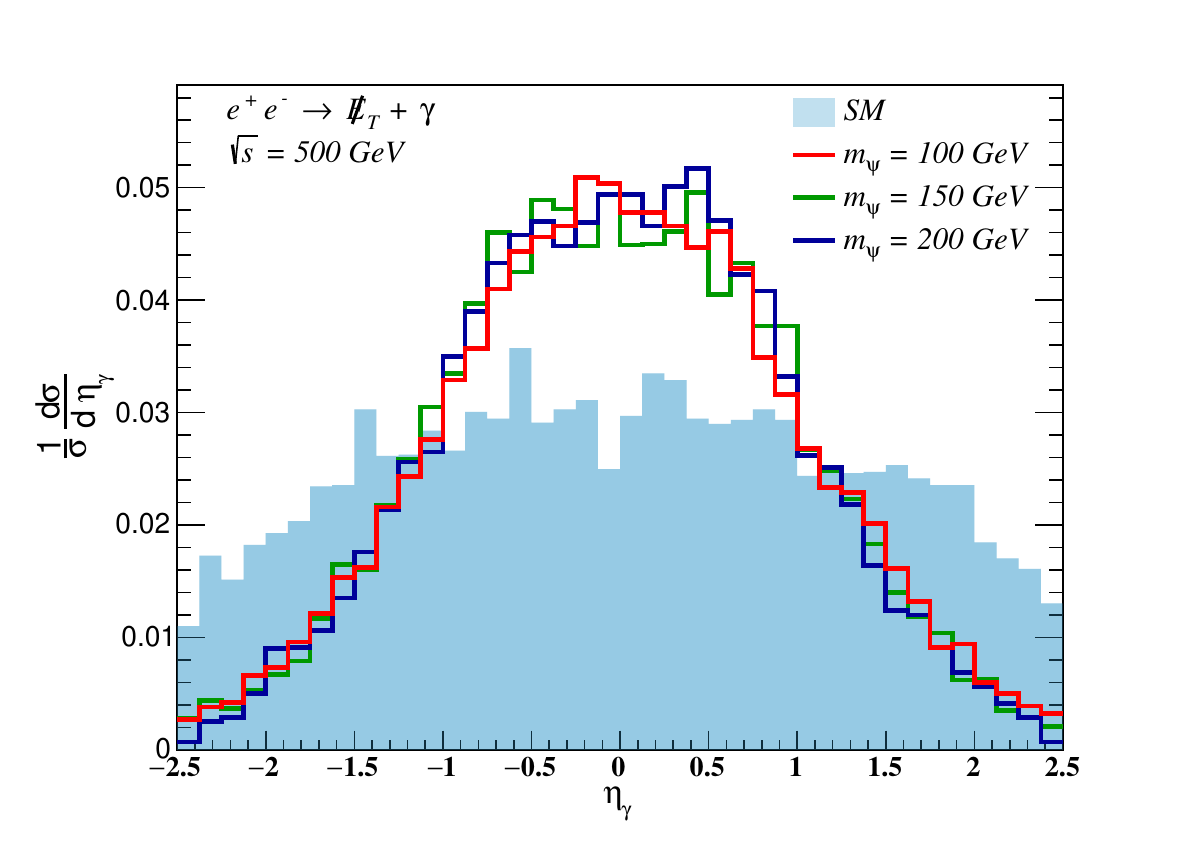}
  \caption{}
  \label{DBfig6}
\end{subfigure}
\hfill
\begin{subfigure}{0.32\textwidth}
  \centering
  \includegraphics[width=\linewidth]{FS_eta.pdf}
  \caption{}
  \label{DBfig7}
\end{subfigure}
\begin{subfigure}{0.32\textwidth}
  \centering
  \includegraphics[width=\linewidth]{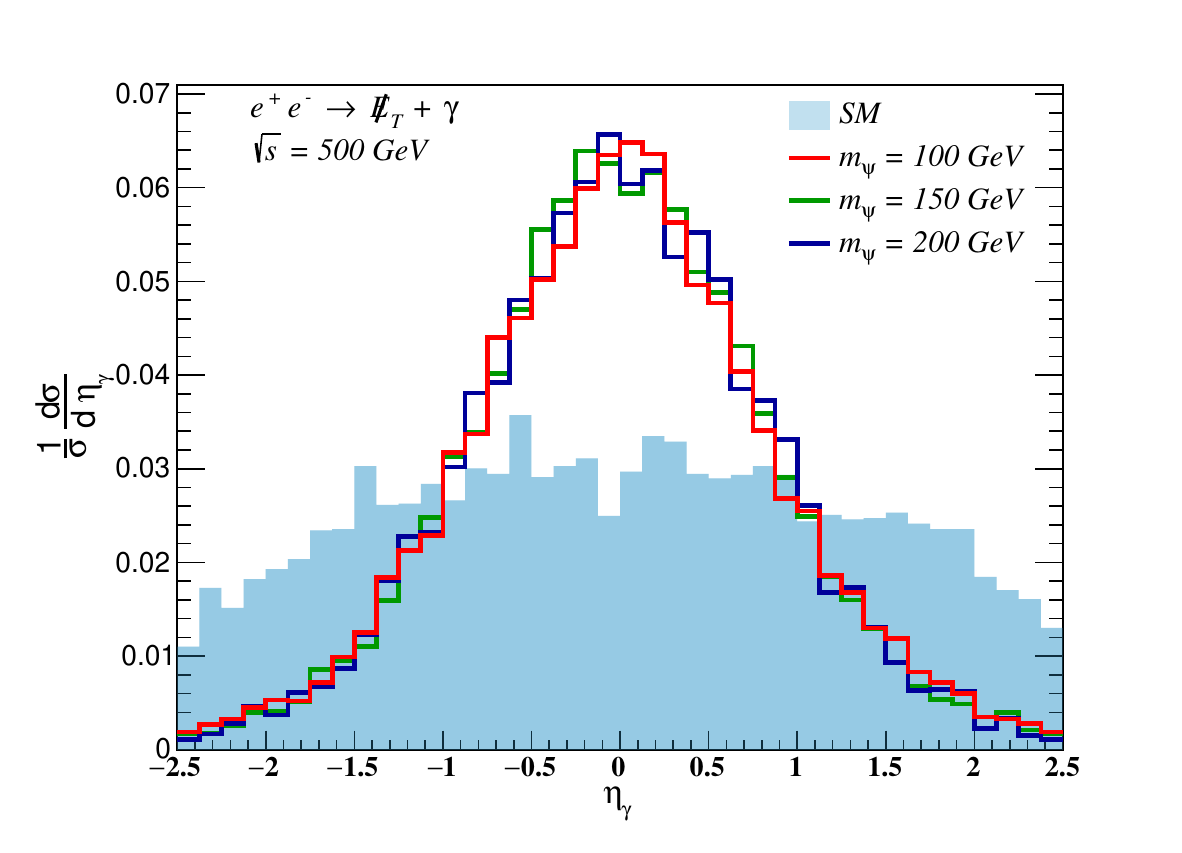}
  \caption{}
  \label{DBfig8}
\end{subfigure}
\caption{Normalized one-dimensional differential cross-sections relative to $p_{T_\gamma}$ (upper panel) and $\eta_{\gamma}$ (lower panel) for both Standard Model processes and those influenced by the $O^\psi_S$, $O^\psi_P$, and $O^\psi_T$ operators, presented at three selected dark matter masses: 100, 150, and 200 GeV.}
\label{fig:DBN_F}
\end{figure}
\begin{figure}[ht]
\centering

\begin{subfigure}{0.45\textwidth}
  \centering
  \includegraphics[width=\linewidth]{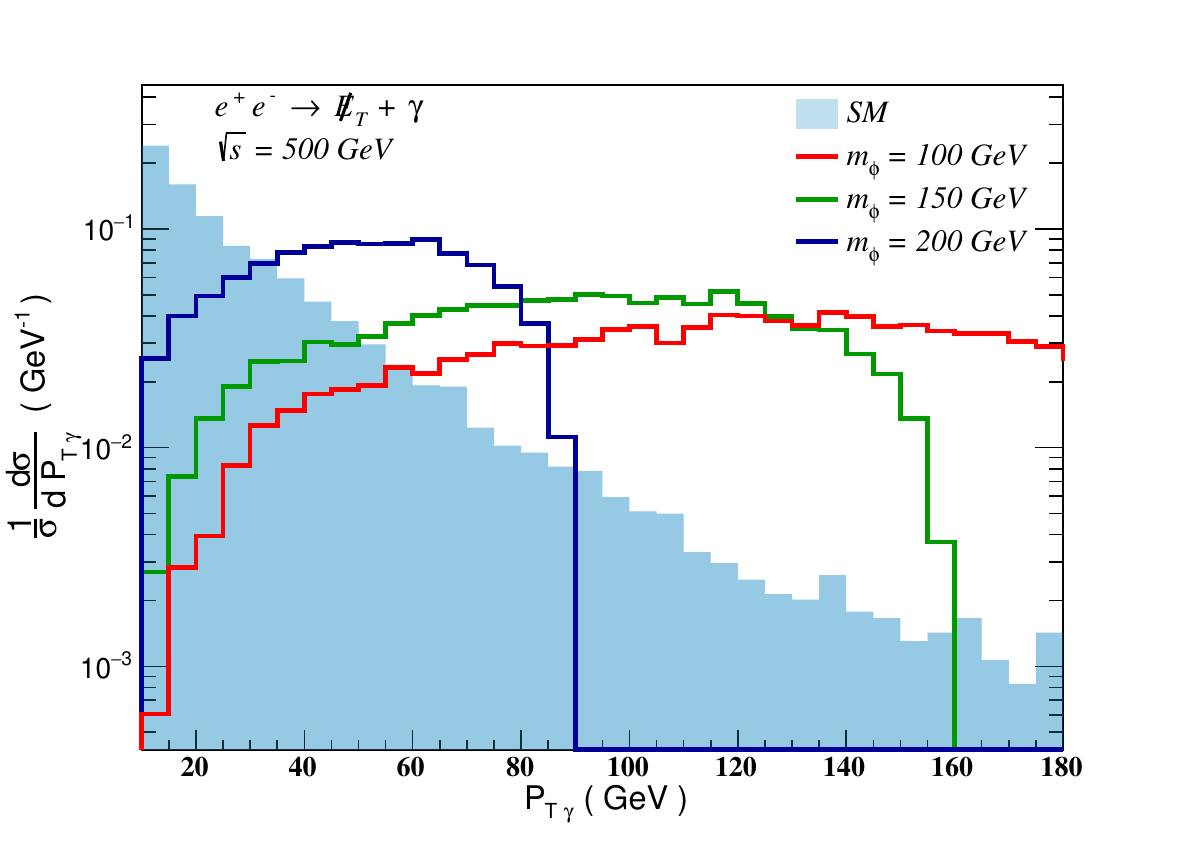}
  \caption{}
  \label{DBfig4}
\end{subfigure}
\hfill
\begin{subfigure}{0.45\textwidth}
  \centering
  \includegraphics[width=\linewidth]{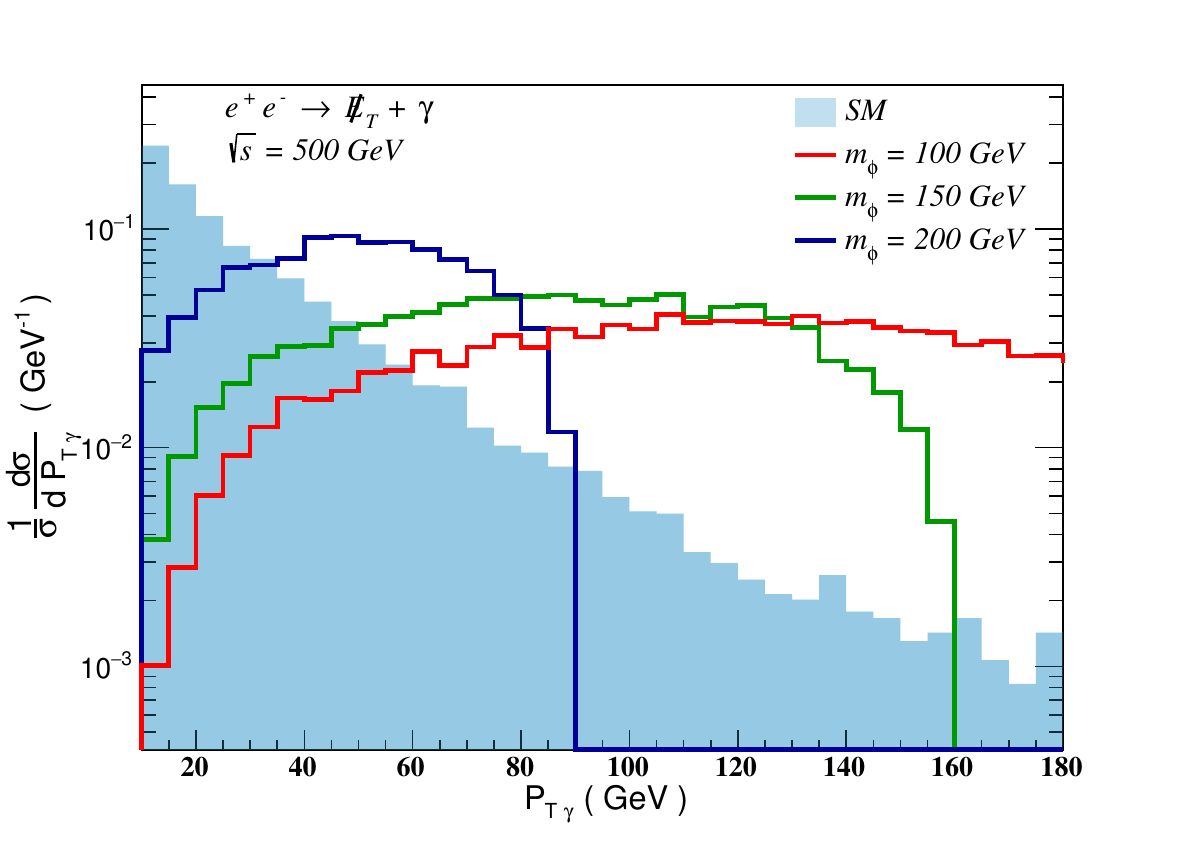}
  \caption{}
  \label{DBfig5}
\end{subfigure}
\begin{subfigure}{0.45\textwidth}
  \centering
  \includegraphics[width=\linewidth]{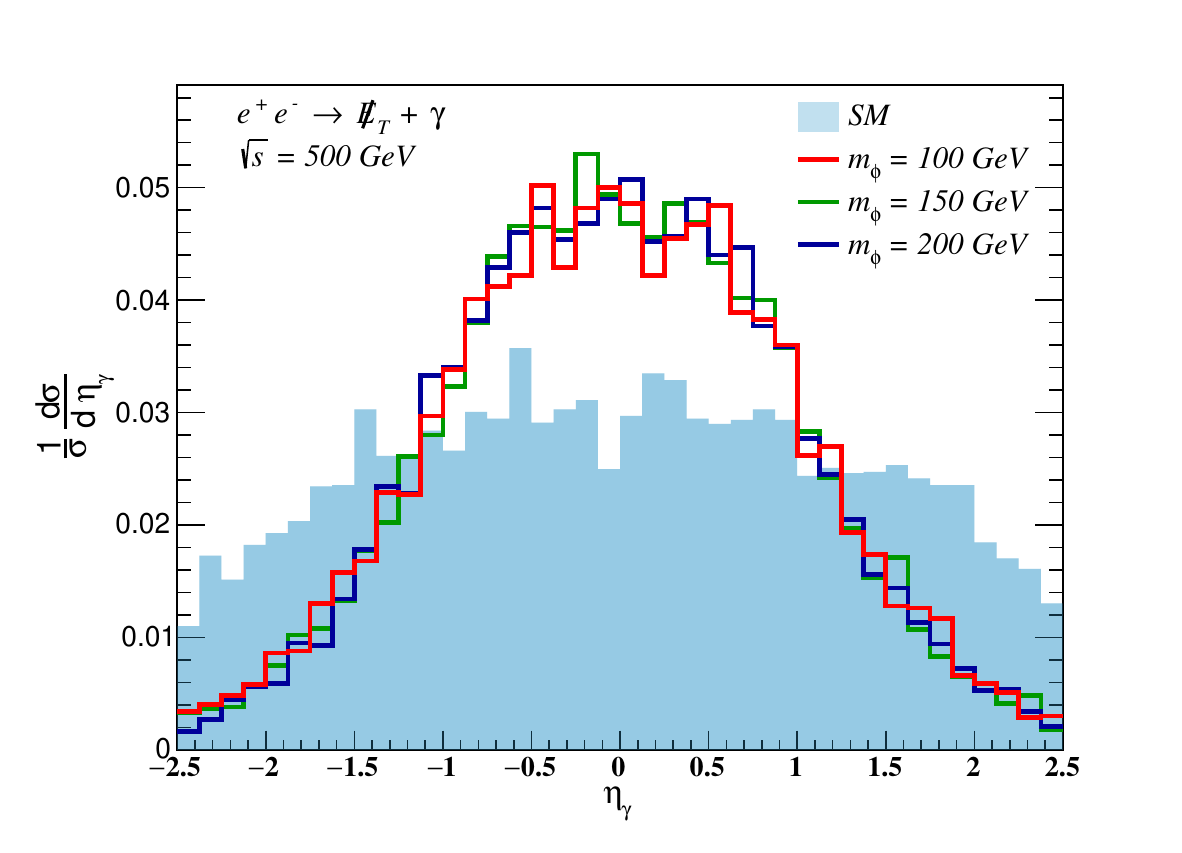}
  \caption{}
  \label{DBfig9}
\end{subfigure}
\hfill
\begin{subfigure}{0.45\textwidth}
  \centering
  \includegraphics[width=\linewidth]{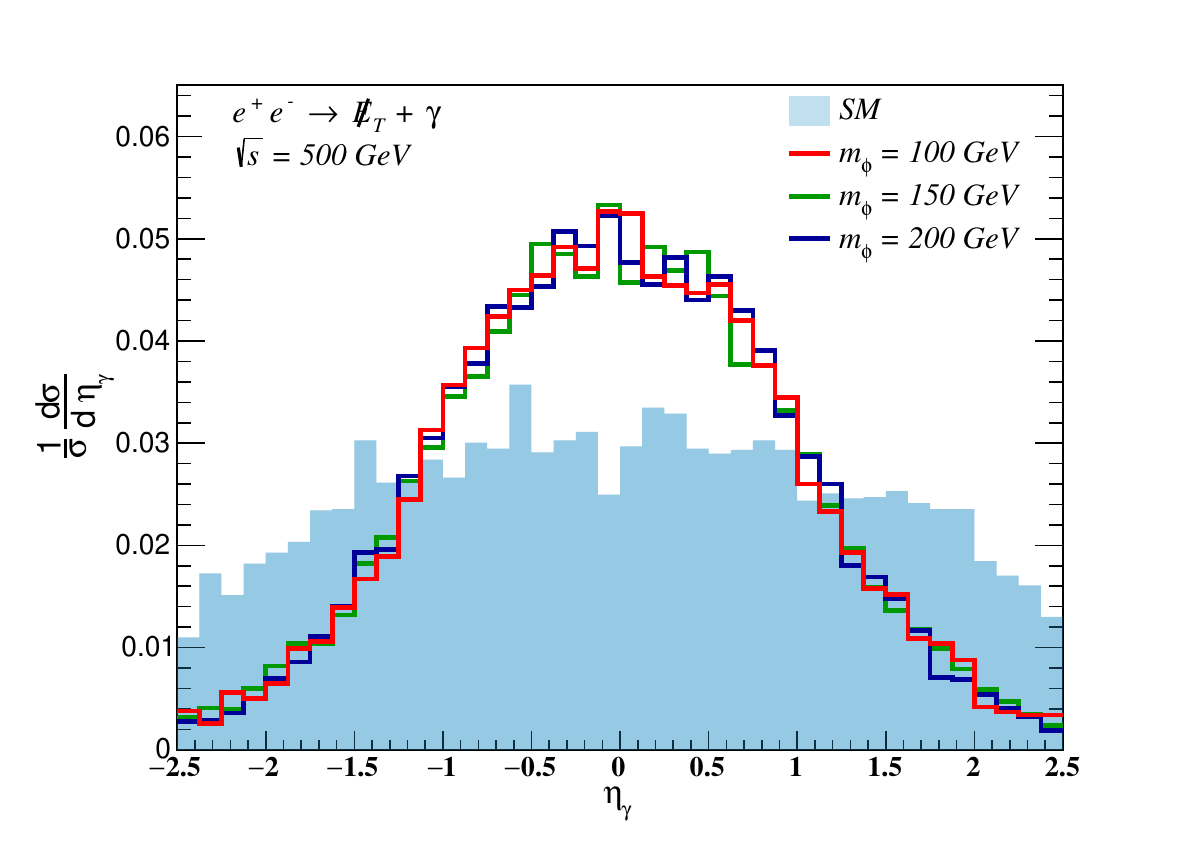}
  \caption{}
  \label{DBfig9}
\end{subfigure}

\caption{Normalized one-dimensional differential cross-sections relative to $p_{T_\gamma}$ (upper panel) and $\eta_{\gamma}$ (lower panel) for both Standard Model processes and those influenced by the $O^\phi_S$ and $O^\phi_T$ operators, presented at three selected dark matter masses: 100, 150, and 200 GeV.}
\label{fig:DBN_S}
\end{figure}

\par We examine the shape profiles associated with processes involving mono-photon with missing energy, focusing on the kinematic observables $p_{{T \gamma}}$ and $\eta_\gamma$ due to their high sensitivity. We generate normalized one-dimensional distributions for both the Standard Model background processes and signals induced by relevant operators. To explore the impact of dark matter (DM) mass, we present the normalized differential cross-sections in figures \ref{DBfig1} - \ref{DBfig4} for three distinct DM mass values, namely $100$, $150$, and $200$ GeV, considering a center-of-mass energy of $\sqrt{s}=500$ GeV and an integrated luminosity of 500 $fb^{-1}$.
 
 The sensitivity of $\Lambda$ to the dark matter mass is heightened through the computation of $\chi^2$ using the double differential distributions of kinematic observables, namely $p_{T_\gamma}$ and $\eta_\gamma$. This analysis is conducted for both background and signal processes under the following conditions:

(i) For a DM mass range of 10 GeV to 125 GeV at a center-of-mass energy ($\sqrt{s}$) of 250 GeV, considering an integrated luminosity of 250 fb$^{-1}$.

(ii) For a DM mass range of 10 GeV to 250 GeV at a center-of-mass energy ($\sqrt{s}$) of 500 GeV, taking into account an integrated luminosity of 500 fb$^{-1}$.
 
The $\chi^2$ is   defined as

\bea{{\chi}}^2  =\sum_{i=1}^{n_1}\sum_{j=1}^{n_2} \left [ \frac{ N_{ij}^{NP}}{\sqrt{  N_{ij}^{SM+NP} +\delta_{\rm sys}^2\ \left({N_{ij}^{SM+NP}}\right)^2}} \right ]^2.
\eea

Here, $ N_{ij}^{NP}$ and $ N_{ij}^{SM+NP}$ denote the new physics and total differential events, respectively, within the two-dimensional grid of ${p_{T_\gamma}}_i$ and ${\eta_\gamma}_j$. The symbol $\delta_{\rm sys}$ represents the overall systematic error associated with the measurement.

\begin{figure}[ht]
\centering

\begin{subfigure}{0.45\textwidth}
  \centering
  \includegraphics[width=\linewidth]{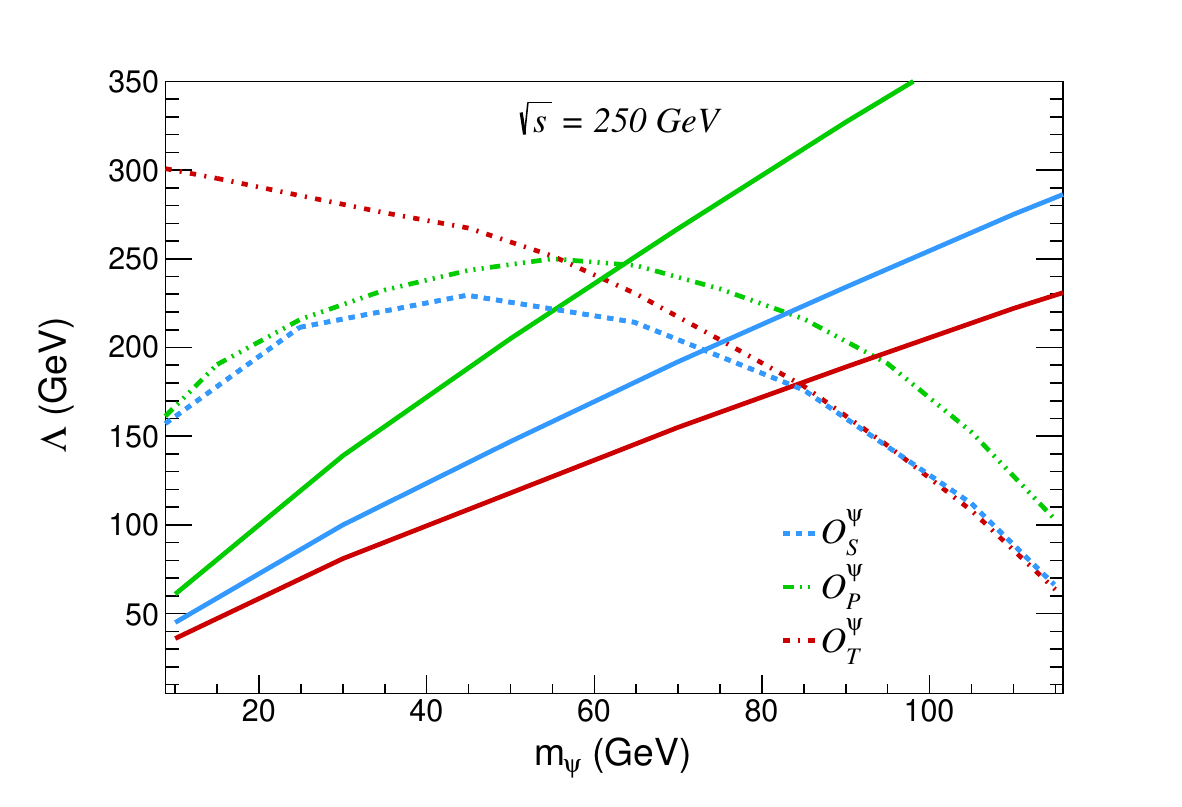}
  \caption{}
  \label{X2F250}
\end{subfigure}
\hfill
\begin{subfigure}{0.45\textwidth}
  \centering
  \includegraphics[width=\linewidth]{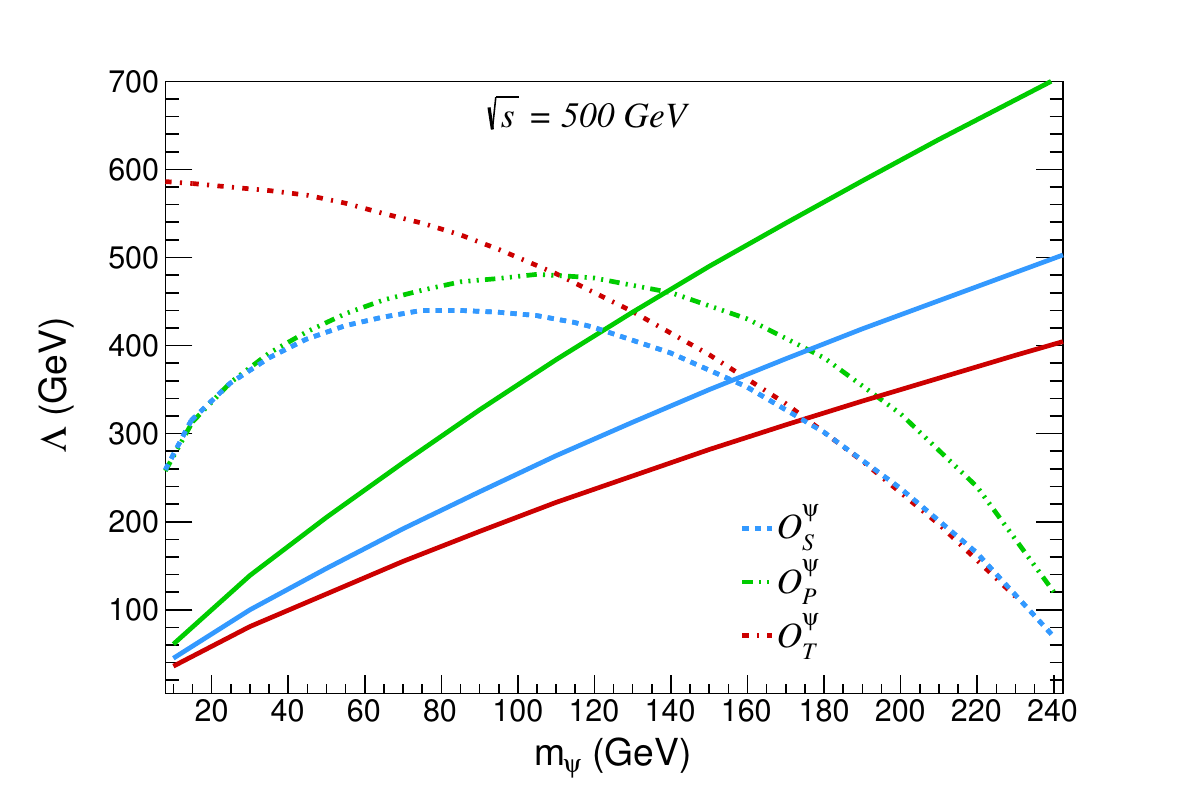}
  \caption{}
  \label{X2F500}
\end{subfigure}
\begin{subfigure}{0.45\textwidth}
  \centering
  \includegraphics[width=\linewidth]{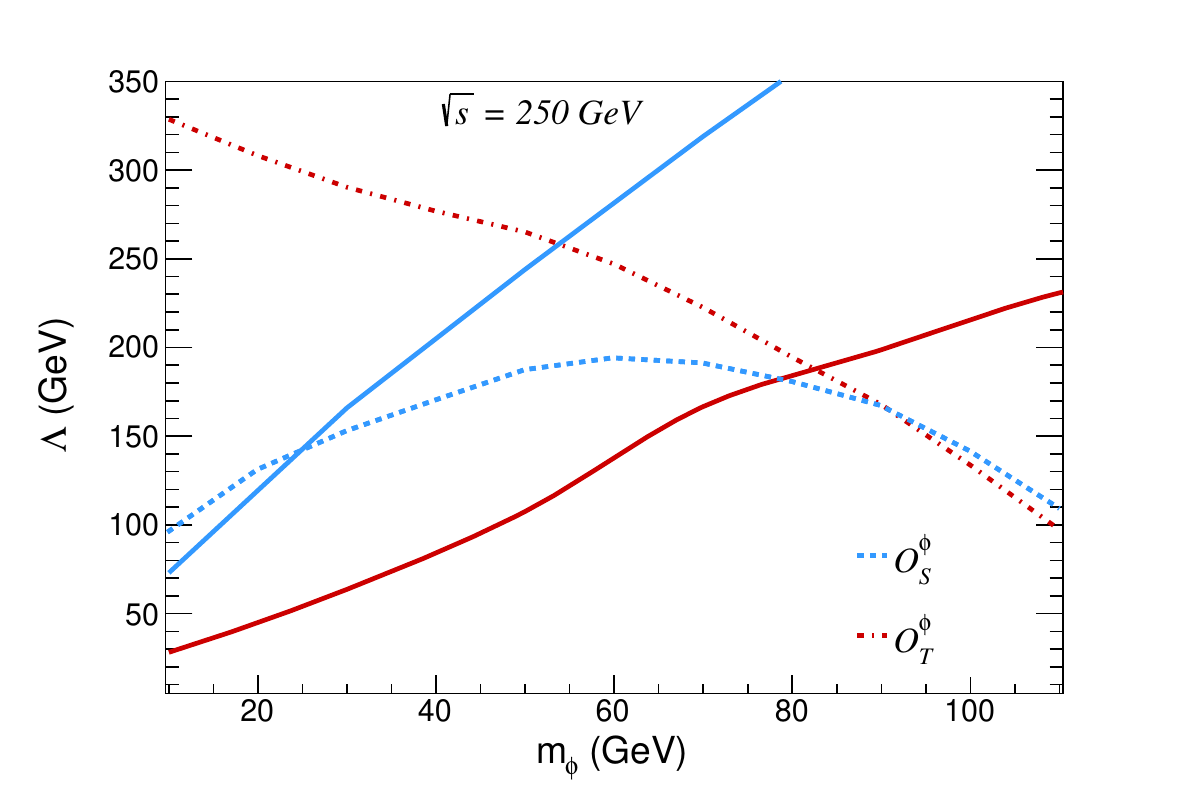}
  \caption{}
  \label{X2S250}
\end{subfigure}
\hfill
\begin{subfigure}{0.45\textwidth}
  \centering
  \includegraphics[width=\linewidth]{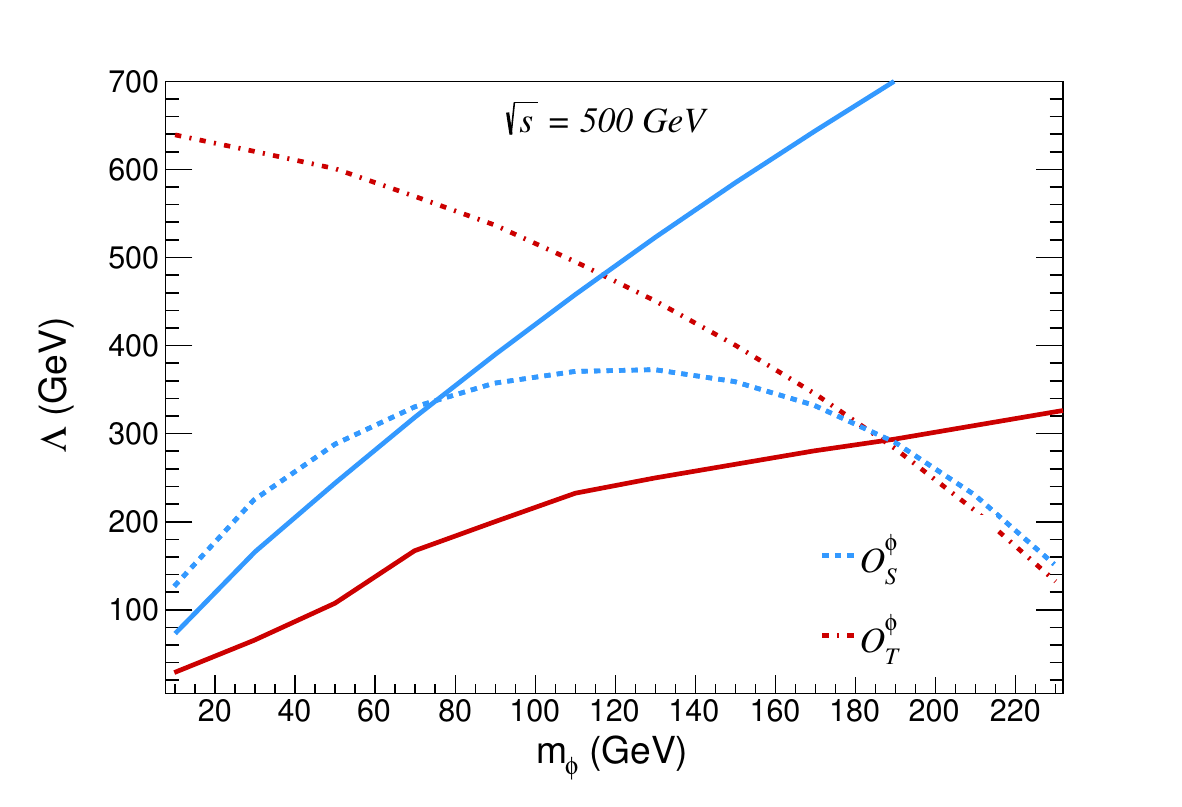}
  \caption{}
  \label{X2S500}
\end{subfigure}

\caption{Dashed lines present $3\sigma$ contours with 99.73\% confidence level in the $m_{\text{DM}}-\Lambda_{\text{eff}}$ plane, derived from $\chi^2$ analyses of the $e^+e^- \to \slash\!\!\! \!E_T +\gamma$ signature at the proposed International Linear Collider (ILC) designed for $\sqrt{s}$ = 250 GeV, with an integrated luminosity of 250 fb$^{-1}$. The region below each dashed line, corresponding to its respective contour, is potentially discoverable with a confidence level of at least 99.73\%. Additionally, the regions below the solid lines corresponding to specific operators satisfy the relic density constraint $\Omega_{\rm DM}h^2 \le 0.1198$.}
\label{fig:ILC_FS}
\end{figure}

 By adopting a conservative estimate of a $1\%$ systematic error and utilizing the collider parameters specified in Table \ref{table:accelparam}, we perform simulations for two-dimensional differential distributions to calculate the ${\chi}^2$. In Figs. \ref{X2F250} to \ref{X2S500}, we present $3\sigma$ contours at the $99.73\%$ confidence level in the $m_{DM}-\Lambda$ plane. These contours correspond to center-of-mass energies of $\sqrt{s}=250$ GeV and 500 GeV, respectively, for the effective operators that adhere to perturbative unitarity.

\section{Conclusion}
This article explores the phenomenology of dark matter within the framework of an effective field theory. The study focuses on SM gauge-invariant contact interactions up to dimension 8 involving dark matter particles and neutral electroweak gauge bosons. To maintain the invariance of SM gauge symmetry above electroweak scale, the investigation is limited to self-conjugate DM particles - specifically, a Majorana fermion and a real scalar boson. The relic density contributions of these particles are estimated, and their parameters are constrained based on the observed relic density of $\Omega_{DM} h^2=0.1198 \pm 0.0012$. Additionally, the annihilation cross-sections for the given DM mass are compared with indirect detection data from H.E.S.S., imposing significant constraints on specific operators, such as $O^{\psi}_{P}$ for fermionic DM and $O^\phi_S$ for scalar DM.

The analysis extends to existing LEP data, revealing that the phenomenologically intriguing DM mass range of $\le 30$ GeV is disallowed, with exceptions for the $O^\psi_P$ and $O^\phi_S$ operators. Subsequently, a $\chi^2$-analysis is conducted for the pair production of DM particles at the proposed ILC, covering a DM mass range of approximately $10-250$ GeV for the operators discussed in the section \ref{model}. The results indicate that within the relic density and indirect detection data-constrained region of the $m_{DM}-\Lambda$ parameter space, enhanced sensitivity can be achieved through the dominant mono-photon signal at the proposed electron-positron collider, ILC.

\section*{Acknowledgments}
We acknowledge the Rajkiya Mahila Mahavidyalaya Budaun Library and its dedicated staff for their invaluable support and resources, which greatly facilitated the completion of this research paper.

\appendix
\section{}
\label{App}

The expressions for thermally averaged cross-section for DM annihilation to photon pair production are calculated to be:
\begin{eqnarray}
\langle \sigma^{\psi}_{S}\ v\rangle\ \left(\psi\ \overline{\psi}\to \gamma \gamma\right)&\simeq&
 \frac{2}{\pi}\ 
\frac{{\alpha^{\psi}_S}^2}{\Lambda^8}\ \cos^4\theta_w\ m_{\psi}^6\ \frac{6}{x_f}\\
\langle\sigma^{\psi}_{P}\  v\rangle\ \left(\psi\ \overline{\psi}\to \gamma \gamma\right)&\simeq&
 \frac{2}{\pi}\ 
\frac{{\alpha^{\psi}_P}^2}{\Lambda^8}\ \cos^4\theta_w\ m_{\psi}^6\  \left(1- \frac{ 3}{2\ x_f}\right)\\
 \langle\sigma^{\psi}_{T}\  v\rangle\ \left(\psi\ \overline{\psi}\to \gamma \gamma\right)&\simeq&
 \frac{2}{3 \pi}\ 
\frac{{\alpha^{\psi}_T}^2}{\Lambda^8}\ \cos^4\theta_w\ m_{\psi}^6\ \frac{6}{x_f}\\
 \langle\sigma^{\phi}_{S}\ v\rangle\ \left(\phi\ {\phi}\to \gamma \gamma\right)&\simeq&
 \frac{16}{\pi}\ 
\frac{{\alpha^{\phi}_S}^2}{\Lambda^8}\ \cos^4\theta_w\ m_{\phi}^6\  \left(1- \frac{ 15}{x_f^2}\right)\\
 \langle\sigma^{\phi}_{T}\  v\rangle\ \left(\phi\ {\phi}\to \gamma \gamma\right)&\simeq&
 \frac{4}{15 \pi}\ 
\frac{{\alpha^{\phi}_T}^2}{\Lambda^8}\ \cos^4\theta_w\ m_{\phi}^6\ \frac{60}{x_f^2}
\end{eqnarray}
\par The expressions for thermally averaged cross-section for DM annihilation to $Z$ pair production are calculated to be:
\begin{eqnarray}
\langle \sigma^{\psi}_{S}\ v\rangle\ \left(\psi\ \overline{\psi}\to Z Z\right)&\simeq&
 \frac{1}{\pi}\ 
\frac{{\alpha^{\psi}_S}^2}{\Lambda^8}\ \sin^4\theta_w\ m_{\psi}^6\ \frac{48}{x_f}\\
\langle\sigma^{\psi}_{P}\  v\rangle\ \left(\psi\ \overline{\psi}\to Z Z\right)&\simeq&
 \frac{2}{\pi}\ 
\frac{{\alpha^{\psi}_P}^2}{\Lambda^8}\ \sin^4\theta_w\ m_{\psi}^6\  \left(8+ \frac{ 6}{ x_f}\right)\\
 \langle\sigma^{\psi}_{T}\  v\rangle\ \left(\psi\ \overline{\psi}\to Z Z\right)&\simeq&
 \frac{2}{3 \pi}\ 
\frac{{\alpha^{\psi}_T}^2}{\Lambda^8}\ \sin^4\theta_w\ m_{\psi}^6\ \frac{6}{x_f}\\
 \langle\sigma^{\phi}_{S}\ v\rangle\ \left(\phi\ {\phi}\to Z Z\right)&\simeq&
 \frac{32}{\pi}\ 
\frac{{\alpha^{\phi}_S}^2}{\Lambda^8}\ \sin^4\theta_w\ m_{\phi}^6\  \left[2+ \left(\frac{m_\phi}{m_Z}\right)^4\ \frac{ 3}{x_f}\right]\\
 \langle\sigma^{\phi}_{T}\  v\rangle\ \left(\phi\ {\phi}\to Z Z\right)&\simeq&
 \frac{3}{2 \pi}\ 
\frac{{\alpha^{\phi}_T}^2}{\Lambda^8}\ \sin^4\theta_w\ m_{\phi}^2\ m_Z^4\ \left(1+\frac{1}{x_f}\right)
\end{eqnarray}
\par The expressions for thermally averaged cross-section for DM annihilation to $Z\ \gamma$ production are calculated to be:
\begin{eqnarray}
\langle \sigma^{\psi}_{S}\ v\rangle\ \left(\psi\ \overline{\psi}\to Z \gamma\right)&\simeq&
 \frac{3}{\pi}\ 
\frac{{\alpha^{\psi}_S}^2}{\Lambda^8}\ \sin^2\theta_w\ \cos^2\theta_w\ m_{\psi}^6\ \frac{6}{x_f}\\
\langle\sigma^{\psi}_{P}\  v\rangle\ \left(\psi\ \overline{\psi}\to Z \gamma\right)&\simeq&
 \frac{4}{\pi}\ 
\frac{{\alpha^{\psi}_P}^2}{\Lambda^8}\ \sin^2\theta_w\ \cos^2\theta_w\  m_{\psi}^6\  \left(2+ \frac{ 3}{ x_f}\right)\\
 \langle\sigma^{\psi}_{T}\  v\rangle\ \left(\psi\ \overline{\psi}\to Z \gamma\right)&\simeq&
 \frac{4}{ \pi}\ 
\frac{{\alpha^{\psi}_T}^2}{\Lambda^8}\ \sin^2\theta_w\ \cos^2\theta_w\  m_{\psi}^6\ \frac{1}{x_f}\\
 \langle\sigma^{\phi}_{S}\ v\rangle\ \left(\phi\ {\phi}\to Z \gamma\right)&\simeq&
 \frac{12}{\pi}\ 
\frac{{\alpha^{\phi}_S}^2}{\Lambda^8}\ \sin^2\theta_w\ \cos^2\theta_w\  m_{\phi}^6\  \left(2- \frac{ m_Z^2}{m_\phi^2}\right)\\
 \langle\sigma^{\phi}_{T}\  v\rangle\ \left(\phi\ {\phi}\to Z \gamma\right)&\simeq&
 \frac{1}{4 \pi}\ 
\frac{{\alpha^{\phi}_T}^2}{\Lambda^8}\ \sin^2\theta_w\ \cos^2\theta_w\  m_{\phi}^2\ m_Z^4\ \left(1+\frac{1}{x_f}\right)
\end{eqnarray}

\end{document}